\documentstyle[12pt,aps]{revtex}
\begin{document}
\title{
\hfill\parbox[t]{2.5in}{\rm\small\baselineskip 14pt
{JLAB-THY-99-01~(revised)}\vfill~}
\vskip 4cm
Beyond the Adiabatic Approximation: \\
\rm{the impact of thresholds on the hadronic spectrum} \\ ~}

\author{Nathan Isgur}
\address{Jefferson Lab, 12000 Jefferson Avenue,
Newport News, Virginia, 23606}
\maketitle

\vspace{2.0 cm}
\begin{center}  {\bf Abstract}  \end{center}

\begin{abstract}

		In the adiabatic approximation, most of the effects of $q \bar q$ loops on spectroscopy can be absorbed into  
a static interquark potential.  I develop a formalism which can be used to treat
the residual nonadiabatic effects associated
with the presence of nearby hadronic thresholds  for heavy quarks.
I then define a  potential which includes additional
high energy corrections to the adiabatic limit which would be present
for finite quark masses. This ``improved" potential
allows a systematic
low energy expansion of the impact of thresholds on hadronic spectra.

\bigskip\bigskip\bigskip\bigskip

\end{abstract}
\pacs{}
\newpage

\section {Introduction}

	    The valence quark model is surprisingly successful at describing mesons and baryons as $q \bar q$
and $qqq$ systems moving in effective potentials.  The surprise comes in part because hadrons are so strongly 
coupled to their (real and virtual) decay channels that each nearby channel  ought to shift a 
hadron's mass by $\Delta m \sim \Gamma_{typical}$, thereby totally disrupting the valence quark model's spectroscopy.

     A simple  resolution of this conundrum has been proposed in a series of papers \cite{GIonV,GIonOZI} 
examining the effects of ``unquenching the quark model", 
{\it i.e.}, allowing extra $q \bar q$ pairs to bubble up in valence quark states.  This
bubbling dresses the valence hadrons with a certain class of meson loop diagrams \cite{Zcomment}.  (These papers also
address how the OZI rule \cite{OZI} survives unquenching; in this paper I will exclusively consider
flavor nonsinglet states for which such OZI-violation is not an issue.) The proposed resolution
is an extension of the idea  \cite{adiabatic} that in the absence of light quarks the heavy 
quarkonium potential $V^{adiabatic}_0(r) \sim b_0r$ is the adiabatically evolving ground state energy 
$E_0(r)$ of the purely
gluonic QCD Hamiltonian in the presence of a static color triplet source $Q$ and color anti-triplet 
sink $\bar Q$ separated by a distance $r$  .  Once $n_f$ light quarks are introduced into this Hamiltonian, two 
major changes occur:
\bigskip 

\begin{enumerate}

\item{  $E_0(r)$ will be shifted to $E_{n_f}(r)$ by ordinary second order perturbation theory, and }

\item{  $E_{n_f}(r)$ will no longer be isolated from all other adiabatic surfaces:  once pair creation can occur,
the $Q \bar Q$ flux tube can break to create states $(Q \bar q)_{\alpha} (q \bar Q)_{\beta}$ with
adiabatic energy surfaces that are constant in $r$ at the values 
$\epsilon_{\alpha}+\epsilon_{\beta}$ ($\epsilon_i$ is the $i^{th}$
eigenvalue of the $Q \bar q$ system, with the heavy quark mass $m_Q$ subtracted).}

\end{enumerate}

\bigskip 
\noindent Despite the latter complication, 
in the weak pair creation limit the flux-tube-like adiabatic surface $E_{n_f}(r)$ can be tracked through
the level crossings that occur when 
$E_{n_f}(r)=\epsilon_{\alpha}+\epsilon_{\beta}$ and {\it identified as the renormalized
$Q \bar Q$ adiabatic potential $V^{adiabatic}_{n_f}(r)=E_{n_f}(r)$}.  In Ref. \cite{GIonV} it is shown that for large $r$,  
$V^{adiabatic}_{n_f}(r)$ remains linear, so that the net effect of the pairs is
simply to renormalize the string tension.  
Since quark modellers determined their string tension from
experiment, {\it the quark model heavy quarkonium 
potential already included the effect of meson loops to leading order in the
adiabatic approximation, i.e., $b_{n_f}=b$, the physical string tension}.

   Note that a  similar renormalization occurs at short distances:  in lowest order  	  
 $\alpha_s^{(0)} \rightarrow \alpha_s^{(n_f)}={{12 \pi} / [{(33-2n_f)ln(Q^2/{\Lambda}_{QCD}^2})}]$.
The renormalization of the string tension by $q \bar q$ loops is quite similar, though complicated by the
existence of the open channels corresponding to adiabatic level crossings. It should also be stressed that the
possibility of subsuming $q \bar q$ loops into $b_{n_f}$ only occurs if one sums over a huge set of
hadronic loop diagrams (real and virtual) \cite{GIonV}.  No simple truncation of the sum over loops, as is often
attempted in hadronic effective theories, is generally possible.  Consider, for example, the simplest orbital
splitting $a_2(1320)-\rho(770)$.  Summing the mass shifts associated with the known decay modes of these
states would significantly change their absolute masses and violently alter their splitting.  Preserving them
requires a large renormalization of the string tension and summing over loop graphs involving many high mass
({\it i.e.}, virtual) channels, since $q \bar q$ creation inside the original $Q \bar Q$ state is dual
to a very large tower of $(Q \bar q)_\alpha (q \bar Q)_\beta$ intermediate states.

    Although the renormalization  $V^{adiabatic}_0 \rightarrow V^{adiabatic}_{n_f}$ 
will capture the bulk of the effect
of ``unquenching" in heavy quarkonia,  $E_{n_f}(r)$ deviates quite substantially from 
linearity near level crossings \cite{GIonV} .  Both this fact and 
explicit modelling suggest that for
phenomenologically relevant quark masses substantial nonadiabatic
effects will remain after renormalization, and in particular 
that states near thresholds to which they are strongly coupled should be expected
to deviate from their potential model positions.  This paper is devoted to 
developing a method for addressing these residual effects. 
This is straightforward as $m_Q \rightarrow \infty$, but I will show that for finite $m_Q$
it is essential to go beyond the naive adiabatic approximation to define
an ``improved" interquark potential which
includes the high energy part of the corrections to the adiabatic limit.

\bigskip
\section {The Formalism  in the Adiabatic Limit}

     To deal with violations of the adiabatic approximation, we can closely imitate the normal methods of mass
renormalization.  For very massive quarks $Q$ and $\bar Q$, the effects of all hadronic loop graphs can be subsumed
into
\begin{equation}
V^{adiabatic}_{n_f}(r)=V^{adiabatic}_0(r)+ \sum_{\alpha \beta}\Delta V^{adiabatic}_{\alpha \beta }(r)
   \label{eq:defineV}
\end{equation}
where $V^{adiabatic}_0(r)$ is the ``purely gluonic" static $Q \bar Q$ potential, 
and $\Delta V^{adiabatic}_{\alpha \beta }(r)$ is
the shift in this 
static potential generated by the channel $\alpha \beta$.  Here the subscript on $V^{adiabatic}_0$ is used to 
denote that it {\it is} purely gluonic; we have suppressed additional labels to identify which gluonic
adiabatic surface $V^{adiabatic}_0$ represents (the normal meson surface, the first $\Lambda = \pm 1$ hybrid surface,
{\it etc.}) since our discussion applies identically to them all.  For the low-lying thresholds of interest to us
here, $\Delta V^{adiabatic}_{\alpha \beta }(r)$ will typically have a strength of order $\Lambda_{QCD}$ and a range of order
$\Lambda_{QCD}^{-1}$.  This range arises because $\psi_{\alpha}(\vec r_{\bar q Q })$ and
$\psi_{\beta}(\vec r_{ q \bar Q})$ are localized at relatively small $\vert \vec r_{\bar q Q } \vert$ 
and $\vert \vec r_{ q \bar Q} \vert$ 
for low-lying states so that for large $r$ the production of such states by the point-like creation of a 
$q \bar q$ pair is strongly damped by the rapidly falling tails of their confined wavefunctions; conversely, for small
$r$ the created $q$ and $\bar q$ are easily accommodated into the ``heart" of their respective wavefunctions.

    Let us now compare the adiabatic Hamiltonian for the $Q \bar Q$ system \cite{Q1Q2}
\begin{equation}
H_{adiabatic}={p^2 \over 2\mu_{Q \bar Q}}+V^{adiabatic}_{n_f}
  \label{eq:H}
\end{equation}
(with $\mu_{ij}$ the reduced mass of $m_i$ and $m_j$)
with the two channel Hamiltonian $H^{(\alpha \beta )}$ that is the penultimate step in generating  
$H_{adiabatic}$ in the sense that all channels {\it except} $\alpha \beta$ have been integrated out:
\bigskip

\begin{equation}
  H^{(\alpha \beta )} = 
  \left[ \matrix{
  {p^2 \over 2\mu_{Q \bar Q}}+
   V^{adiabatic}_{n_f} -\Delta V^{adiabatic}_{\alpha \beta }  & H_{(\alpha \beta )}^{q \bar q} & \cr
   H_{(\alpha \beta )}^{q \bar q}                            & {p^2_{\rho} \over 2\mu_{\alpha \beta}}+\epsilon_{\alpha}+\epsilon_{\beta}& \cr
    }\right]
   \label{eq:Hmatrix}
\end{equation}
\vskip 0.5cm
\noindent where $H_{(\alpha \beta )}^{q \bar q}$ is an interaction which couples the $Q \bar Q$ system to {\it the single
channel} $(Q \bar q)_\alpha (q \bar Q)_\beta$ with the matrix elements dictated by the underlying pair creation 
Hamiltonian $H_{pc}^{q \bar q}$.  In the adiabatic limit we must recover $H_{adiabatic}$ from $H^{(\alpha \beta)}$, but 
$H^{(\alpha \beta )}$ contains the full dynamics of the coupling of the $Q \bar Q$ system to the
channel $\alpha \beta$.  With the superscript on $H^{(\alpha \beta)}$ we are making explicit that $H^{(\alpha \beta)}$
has the channel  $\alpha \beta$ removed from the $Q \bar Q$ adiabatic potential and added back in full
via $H ^{q \bar q} _{(\alpha \beta)}$.  We could in general remove any subset of $n$ channels from 
$V^{adiabatic}_{n_f}(r)$ and add
them back in dynamically as part of an $(n + 1)-$channel problem.  In the limit of taking all channels we
would recover the original full unquenched Hamiltonian.  However, 
since our treatment is in lowest-order perturbation theory,
the effects of the individual channels are additive, and Eq.~(\ref{eq:Hmatrix}) with 
just an individual channel $(\alpha \beta)$
selected for study is sufficient for our purposes.

   Note that the hadronic multichannel version of our unquenched Hamiltonian is 
an appropriate representation
of $q \bar q$ pair creation in a confined system.  When the pair is created, the $(Q \bar q q \bar Q)$ system has
three relative coordinates which we may take to be $\vec \rho$, the separation between the center of mass of meson
$\beta$ and that of meson $\alpha$, and the two intrameson coordinates  
$\vec r_{\bar q Q } \equiv \vec r_{\bar q}-\vec r_{Q }$ and
$\vec r_{q \bar Q }  \equiv \vec r_{q}-\vec r_{ \bar Q }$.
Since we ignore the residual final state interaction between the color singlets $(Q \bar q)_{\alpha}$ and
$(q \bar Q)_{\beta}$, the eigenstates of this sector are  
mesons in relative plane waves, corresponding to the entry $H^{(\alpha \beta )}_{22}$
in Eq.~(\ref{eq:Hmatrix}). Thus, with $\vec p_{\rho}$ canonically conjugate to $\vec \rho$,
the three quantum labels $(\vec p_{\rho}, \alpha, \beta)$ replace the three labels $(\vec \rho, 
\vec r_{ \bar q Q}, \vec r_{ q \bar Q})$.   

   The main goal of this paper is to describe the relation between 
the eigenvalues of the adiabatic Hamiltonian  (\ref{eq:H})
and the dynamic Hamiltonian (\ref{eq:Hmatrix}).  If we define
\bigskip

\begin{equation}
  H_0 = 
  \left[ \matrix{
  {p^2 \over 2\mu_{Q \bar Q}}+V^{adiabatic}_{n_f}   &            0                                                         & \cr
  0                               & {p^2_{\rho} \over 2\mu_{\alpha \beta}}+\epsilon_{\alpha}+\epsilon_{\beta}   & \cr
    }\right]
\end{equation}
and
\begin{equation}
  H_{pert} = 
  \left[ \matrix{
  -\Delta V^{adiabatic}_{\alpha \beta }       & H_{(\alpha \beta )}^{q \bar q} & \cr
   H_{(\alpha \beta )}^{q \bar q}   & 0                              & \cr
    }\right]   
\end{equation}

\vskip 0.5cm
\noindent and denote  the $Q \bar Q$ eigenvalues of $H_0$ and 
$H ^{(\alpha \beta)}$ by $E^0_i$ and $E^{(\alpha \beta )}_i$,
respectively, then since $H ^{(\alpha \beta)}=H_0+H_{pert}$, by second order perturbation theory  
$\Delta E^{(\alpha \beta )}_i=E^{(\alpha \beta )}_i-E^0_i$ is given by
\begin{eqnarray}
\Delta E_i^{(\alpha \beta )}
 &=&
-\langle \psi^i_0 \vert \Delta V^{adiabatic}_{\alpha \beta } \vert \psi^i_0  \rangle
+\int d^3q 
{
{\vert \langle \alpha \beta (\vec q) \vert H_{(\alpha \beta )}^{q \bar q} \vert \psi^i_0  \rangle \vert^2}
\over
{E_i^0-(\epsilon_{\alpha}+\epsilon_{\beta}+{q^2 \over {2\mu_{\alpha \beta}}})}
}
\\
&\equiv& -\Delta E_i^{adiabatic(\alpha \beta)}+\Delta E_i^{dynamic(\alpha \beta)}~,
   \label{eq:DeltaE}
\end{eqnarray}
where $\vert \psi^i_0 \rangle$ is the $i^{th}$ eigenstate of $H_0$. {\it This simple equation 
is the main focus of this paper.} It represents the correction to the adiabatic
approximation for the $Q \bar Q$ energy eigenvalues from a full dynamical versus  an adiabatic 
treatment of the channel $(\alpha \beta)$. In what follows I will first show
explicitly that $\Delta E_i^{(\alpha \beta )} \rightarrow 0$ as expected in the limit
$m_Q \rightarrow \infty$. I will then define an improved effective potential  $V^{improved}_{n_f}$
which incorporates ``trivial" high energy corrections
to the adiabatic approximation, but which is essential for incorporating threshold effects 
in a systematic low energy expansion for finite $m_Q$.

   I begin by defining precisely  $\Delta V^{adiabatic}_{\alpha \beta }$
in Eq.~(\ref{eq:defineV}).  If ${\vert \alpha \beta (\vec \rho) \rangle}$ denotes an $\alpha \beta$ state with relative
coordinate $\vec \rho$, then as $m_Q \rightarrow \infty$

\begin{equation}
\langle \alpha \beta (\vec \rho) \vert H_{pc}^{q \bar q} \vert Q \bar Q (\vec r)  \rangle
\equiv \langle \alpha \beta (\vec \rho) \vert H_{\alpha \beta}^{q \bar q} \vert Q \bar Q (\vec r)  \rangle
=c_{\alpha \beta} (\vec r) \delta^3 (\vec \rho - \vec r)
 \label{eq:defc}
\end{equation}
since the $Q \bar Q$ relative coordinate is frozen in the adiabatic approximation by definition
and
since $\vec r_{\bar Q}-\vec r_{Q} \rightarrow \vec \rho$
as $m_Q \rightarrow \infty$. Thus

\begin{eqnarray}
\langle Q \bar Q (\vec r~')  \vert \Delta V^{adiabatic}_{\alpha \beta } \vert Q \bar Q (\vec r)  \rangle
&\equiv&
\int d^3\rho 
{
{\langle Q \bar Q (\vec r~') \vert H_{\alpha \beta}^{q \bar q} \vert \alpha \beta (\vec \rho) \rangle 
\langle \alpha \beta (\vec \rho) \vert H_{\alpha \beta}^{q \bar q} \vert Q \bar Q (\vec r) \rangle}
\over
{br-(\epsilon_{\alpha}+\epsilon_{\beta})}
} \label{eq:DeltaVadiabatic} \\
& = & \delta^3(\vec r~' - \vec r) 
{
{\vert c_{\alpha \beta}(\vec r) \vert^2}
\over
{br-(\epsilon_{\alpha}+\epsilon_{\beta})}
}     
\label{eq:DeltaVad}  \\
& \equiv & \delta^3(\vec r~' - \vec r)\Delta V^{adiabatic}_{\alpha \beta }(\vec r)
\end{eqnarray}
for $br$ far from $\epsilon_{\alpha}+\epsilon_{\beta}$; see
Ref. \cite{GIonV} for a discussion of how the poles in (\ref{eq:DeltaVad}) 
are to be handled.
Given this $\Delta V^{adiabatic}_{\alpha \beta }(\vec r)$, by definition
\begin{eqnarray}
\Delta E_i^{adiabatic(\alpha \beta)}&=& \langle \psi_0^i \vert \Delta V^{adiabatic}_{\alpha \beta }  \vert \psi_0^i \rangle \\
&=& \int d^3r 
{
{\vert \psi_0^i(\vec r) \vert^2  \vert c_{\alpha \beta}(\vec r) \vert^2 }
\over
{br-(\epsilon_{\alpha}+\epsilon_{\beta})}
} ~. 
\label{eq:DeltaEad}
\end{eqnarray}

     I now show how $\Delta E_i^{adiabatic(\alpha \beta)}$ approximates the true shift
\begin{equation}
\Delta E_i^{dynamic(\alpha \beta)}
\equiv
\int d^3q 
{
{\vert \langle \alpha \beta (\vec q) \vert H_{(\alpha \beta )}^{q \bar q} \vert \psi^i_0  \rangle \vert^2}
\over
{E_i^0-(\epsilon_{\alpha}+\epsilon_{\beta}+{q^2 \over {2\mu_{\alpha \beta}}})}
}
   \label{eq:DeltaEdynamic}
\end{equation}
even for ``nearby" thresholds as $m_Q \rightarrow \infty$. Denote by
$\langle v \rangle_i$  the expectation value of the variable
$v$ in the state $\vert \psi^i_0 \rangle$. In the limit $m_Q \rightarrow \infty$, each of
${ {\langle p^2 \rangle_i} \over 2 \mu_{Q \bar Q}}$, $b\langle r \rangle_i$, and
${ {q^2} \over 2 \mu_{\alpha \beta}}$ vanishes like $({\Lambda_{QCD} \over m_Q})^{1/3}\Lambda_{QCD}$
and so is small compared to $\epsilon_{\alpha}+\epsilon_{\beta}$, which is of order $\Lambda_{QCD}$,
but large compared to the corrections to $\epsilon_{\alpha}+\epsilon_{\beta}$, which are
of order $\Lambda_{QCD}/m_Q$.
(In the general power law potential $c_nr^n$, they each behave like
$({\Lambda_{QCD} \over m_Q})^{n/{n+2}}\Lambda_{QCD}$,
{\it i.e.}, they vanish for any confining ($n>0$) potential).  For
${ {q^2} \over 2 \mu_{\alpha \beta}}$, this statement is nontrivial:  
it relies on the behavior of the numerator of Eq.~(\ref{eq:DeltaEdynamic}).
Using  Eq.~(\ref{eq:defc}), 
\begin{eqnarray}
\langle \alpha \beta (\vec q) \vert H^{q \bar q}_{(\alpha \beta)}\vert Q \bar Q (\vec p) \rangle &=&
{1 \over {(2 \pi)^3}}
\int d^3r e^{i(\vec p-\vec q)\cdot \vec r} c_{\alpha \beta}(\vec r)  \\
&\equiv & \tilde c_{\alpha \beta}(\vec p-\vec q)~,
\end{eqnarray}
so, even though
$\vert \vec p \vert \sim (\Lambda^2_{QCD}m_Q)^{1/3} \rightarrow \infty$,
$\vert \vec p-\vec q \vert$ must be of order $\Lambda_{QCD}$ 
since $\tilde c_{\alpha \beta}$ is a light quark object. 
After writing $E_i^0={{\langle p^2 \rangle_i} \over 2 \mu_{Q \bar Q}}+b\langle r \rangle_i$, 
we can therefore Taylor series expand:
\begin{eqnarray}
\Delta E_i^{dynamic(\alpha \beta)}
&\simeq& -({1 \over {\epsilon_{\alpha}+\epsilon_{\beta}}})
\int d^3q 
\vert \langle \alpha \beta (\vec q) \vert H_{(\alpha \beta )}^{q \bar q} \vert \psi^i_0  \rangle \vert^2 \nonumber \\
&&( 1+({1 \over {\epsilon_{\alpha}+\epsilon_{\beta}}})  
[{{\langle p^2 \rangle_i-q^2} \over {2 \mu_{Q \bar Q}}}+b\langle r \rangle_i]+ \cdot \cdot \cdot ) 
\label{eq:Taylorseries} \\
&\simeq& -({1 \over {\epsilon_{\alpha}+\epsilon_{\beta}}})
\int d^3p' \int d^3q \int d^3p~ 
\phi^{i*}_0(\vec p~') 
\tilde c^*_{\alpha \beta}(\vec p~'-\vec q) 
\tilde c_{\alpha \beta}(\vec p-\vec q)
\phi^{i}_0(\vec p)  \nonumber \\
&&( 1+({1 \over {\epsilon_{\alpha}+\epsilon_{\beta}}}) 
[{{\langle p^2 \rangle_i-q^2} \over {2 \mu_{Q \bar Q}}}+b\langle r \rangle_i]+ \cdot \cdot \cdot )~,
\end{eqnarray}
where $\phi^{i}_0(\vec p) \equiv {1 \over (2 \pi)^{3 \over 2}}\int d^3r e^{-i \vec p \cdot \vec r}\psi^{i}_0(\vec r)$.
Noting that
\begin{equation}
\int d^3s ~\tilde c_{\alpha \beta}(\vec s)= c_{\alpha \beta}(\vec 0)
\end{equation}
and that except for the $\tilde c$ the factors of the integrand  are slowly varying functions,
we can approximate
\begin{equation}
\tilde c_{\alpha \beta}(\vec s) \simeq c_{\alpha \beta}(\vec 0) \delta^3(\vec s)+ \cdot \cdot \cdot 
\end{equation}
to obtain
\begin{eqnarray}
\Delta E_i^{dynamic(\alpha \beta)}
&\simeq& -({\vert c_{\alpha \beta}(\vec 0)\vert^2 \over {\epsilon_{\alpha}+\epsilon_{\beta}}})
\int d^3p 
\vert \phi^i_0(\vec p) \vert^2  
( 1+({1 \over {\epsilon_{\alpha}+\epsilon_{\beta}}})  
[{{\langle p^2 \rangle_i-p^2} \over {m_Q}}+b\langle r \rangle_i]+ \cdot \cdot \cdot ) \\
&\simeq& -({\vert c_{\alpha \beta}(\vec 0)\vert^2 \over {\epsilon_{\alpha}+\epsilon_{\beta}}})
\int d^3p 
\vert \phi^i_0(\vec p) \vert^2  
[ 1+({br \over {\epsilon_{\alpha}+\epsilon_{\beta}}})]  \\
&\simeq& 
\int d^3p 
{{\vert \phi^i_0(\vec p) \vert^2 \vert c_{\alpha \beta}(\vec 0)\vert^2}
\over 
{br - {(\epsilon_{\alpha}+\epsilon_{\beta})}}} \\
&\simeq& 
\int d^3r 
{{\vert \psi^i_0(\vec r) \vert^2 \vert c_{\alpha \beta}(\vec 0)\vert^2}
\over 
{br - {(\epsilon_{\alpha}+\epsilon_{\beta})}}} ~.
\label{eq:penultimateadiabatic}
\end{eqnarray}
This expression differs slightly from 
$\Delta E_i^{adiabatic(\alpha \beta)}$ in Eq.~(\ref{eq:DeltaEad}):  it has
$\vert c_{\alpha \beta}(\vec r)\vert^2 \rightarrow \vert c_{\alpha \beta}(\vec 0)\vert^2$.  
However, for low-lying states, 
$\langle{
{\vert c_{\alpha \beta}(\vec r)\vert^2}
\over
{\vert c_{\alpha \beta}(\vec 0)\vert^2}
}\rangle_i
-1 \sim ({\Lambda_{QCD} \over m_Q} )^{2/3}
$
which is negligible as $m_Q \rightarrow \infty$ compared to 
$\langle{
{br}
\over
{\epsilon_{\alpha}+\epsilon_{\beta}}
}\rangle_i
\sim ({\Lambda_{QCD} \over m_Q} )^{1/3}
$
which we retained.  (The physics behind this approximation is simply that 
$\vert c_{\alpha \beta}(\vec r)\vert^2$
reflects light quark scales
while 
$\vert \psi^i_0(\vec r) \vert^2$
reflects short distance scales as $m_Q \rightarrow \infty$.)  
Thus to leading order  as $m_Q \rightarrow \infty$,
$\Delta E_i^{adiabatic(\alpha \beta)}=\Delta E_i^{dynamic(\alpha \beta)}$
for low-lying states,
as we set
out to prove. 

   The deviation $\Delta E_i^{(\alpha \beta)}$ of the energy
of the state $i$ from its value in the adiabatic potential $V^{adiabatic}_{n_f}$ due
to the residual dynamical effects of the channel $(Q \bar q)_{\alpha} (q \bar Q)_{\beta}$
thus has the property that $\Delta E_i^{(\alpha \beta)} \rightarrow 0$
channel by channel as $m_Q \rightarrow \infty$. The utility of this formalism for heavy quarkonium 
arises from not only this property, but
also that it allows one to consistently focus on low-lying thresholds.
The latter feature is based on the fact that the full shift
$\Delta E_i \equiv \sum_{\alpha \beta} \Delta E_i^{(\alpha \beta)}$
may receive significant ``random" contributions from strategically placed low mass channels,
but for fixed large $m_Q$ the
$\Delta E_i^{(\alpha \beta)} \rightarrow 0$ rapidly as ${\epsilon_{\alpha}+\epsilon_{\beta}}$
gets large. This rapid convergence occurs because, first,  as ${\epsilon_{\alpha}+\epsilon_{\beta}} \rightarrow \infty$,
$E_i^0$ and $q^2/2 \mu_{\alpha \beta}$ in Eq. (\ref{eq:DeltaEdynamic}) and $br$ in Eq. (\ref{eq:DeltaEad}),
which were already small with respect to ${\epsilon_{\alpha}+\epsilon_{\beta}}$
even for low mass channels, become negligible. In this limit $\Delta E_i^{(\alpha \beta)}$
is therefore trivially zero independent of the accuracy of the approximations inherent in 
Eqs. (\ref{eq:Taylorseries})-(\ref{eq:penultimateadiabatic}). Moreover,
the factor $\vert c_{\alpha \beta}(\vec r)\vert^2$ in the numerator of each of
$\Delta E_i^{adiabatic(\alpha \beta)}$ and $\Delta E_i^{dynamic(\alpha \beta)}$ rapidly
approaches zero as ${\epsilon_{\alpha}+\epsilon_{\beta}} \rightarrow \infty$
since for low-lying states $\vert \psi_i^0 \rangle$ there is little kinetic energy
in the initial wavefunction and the pair creation process can only create momenta of order $\Lambda_{QCD}$.
We can therefore expect this formalism to provide a rapidly converging low-energy expansion of the
effects of pair creation on heavy quarkonia.

\bigskip
\section {An Improved Quarkonium Potential}

    While suitable for heavy quarkonia, the framework of Section II has serious shortcomings for light
quark spectroscopy. Because the eigenvalues $\epsilon^{(m_Q)}_\alpha$ and $\epsilon^{(m_Q)}_\beta$ and the matrix elements
$\langle \alpha^{(m_Q)} \beta^{(m_Q)} (\vec \rho ) \vert H_{pc}^{q \bar q} \vert  Q \bar Q (\vec r) \rangle$
for finite $m_Q$ are only qualitatively related to their
$m_Q \rightarrow \infty$ counterparts $\epsilon_\alpha$, $\epsilon_\beta$, and 
$\langle \alpha \beta (\vec \rho ) \vert H_{pc}^{q \bar q} \vert  Q \bar Q (\vec r) \rangle$,
$\Delta V^{adiabatic}_{\alpha \beta}(r)$ is not in this case an accurate approximation to
the effects of the channel $(Q \bar q)_{\alpha} (q \bar Q)_{\beta}$. As a result, the $\Delta E_i^{(\alpha \beta)}$
will not be small, {\it i.e.}, the critical separation of the effects of the channel $\alpha \beta$ into
large adiabatic and small residual dynamical effects will fail. 
If it were only for a few nearby channels, this failure would not be so serious, but while
$E_i^0$ and $q^2/2 \mu_{\alpha \beta}$ in the finite $m_Q$ analogue
of Eq. (\ref{eq:DeltaEdynamic}) and $br$ in Eq. (\ref{eq:DeltaEad}) 
can still be neglected as $\epsilon_\alpha+\epsilon_\beta \rightarrow \infty$, since
$\epsilon^{(m_Q)}_\alpha+\epsilon^{(m_Q)}_\beta \neq \epsilon_\alpha+\epsilon_\beta$
and
$\langle \alpha^{(m_Q)} \beta^{(m_Q)} (\vec \rho ) \vert H_{pc}^{q \bar q} \vert  Q \bar Q (\vec r) \rangle \neq
\langle \alpha \beta (\vec \rho ) \vert H_{pc}^{q \bar q} \vert  Q \bar Q (\vec r) \rangle$,
the finite $m_Q$ analogue of $\Delta E_i^{dynamical(\alpha \beta)}$
will not trivially approach $\Delta E_i^{adiabatic(\alpha \beta)}$
in this high energy limit. Moreover, while the matrix elements
$\langle \alpha^{(m_Q)} \beta^{(m_Q)} (\vec \rho ) \vert H_{pc}^{q \bar q} \vert  Q \bar Q (\vec r) \rangle$
may still be expected to cut off high mass channels, since the
momenta in low mass states and in the pair creation process are comparable, these channels will
be cut off more slowly in the finite $m_Q$ analogue of $\Delta E_i^{dynamical(\alpha \beta)}$
than in $\Delta E_i^{adiabatic(\alpha \beta)}$. These shortcomings make the 
$m_Q \rightarrow \infty$ framework far less useful in light quark systems since
$\Delta E_i \equiv \sum_{\alpha \beta} \Delta E_i^{(\alpha \beta)}$
will converge only marginally faster than the ``brute force" sum
$\sum_{\alpha \beta} \Delta E_i^{dynamic(\alpha \beta)}$.

    I will now show that it is possible to define an improved effective
quarkonium potential $V^{improved}_{n_f}$ which leads to energy shifts
$\delta E_i^{(\alpha \beta)}$ which vanish as $\epsilon_{\alpha}+\epsilon_{\beta} \rightarrow \infty$
{\it for any $m_Q$} and so give smaller and more rapidly converging corrections to the quark model
spectroscopy built on $V^{improved}_{n_f}$ than the $\Delta E_i^{(\alpha \beta)}$. 
The price to be paid for this important feature is that
the universal (flavor-independent) adiabatic quarkonium potential $V^{adiabatic}_{n_f}$ must be
replaced by a flavor-dependent effective potential $V^{improved}_{n_f}$
built out of $V^{adiabatic}_0$ plus flavor-dependent contributions $\Delta V^{improved}_{\alpha \beta}$.

    The basic idea is very simple. For any $m_Q$ \cite{Q1Q2}, the shift
in the energy of the state $\vert \psi^{i~(m_Q)}_0 \rangle$ due to channel $\alpha \beta$ is given by
the generalization of
Eq.~(\ref{eq:DeltaEdynamic}), namely
\begin{equation}
\Delta E_i^{dynamic(\alpha \beta)(m_Q)}
\equiv
\int d^3q 
{
{\vert \langle \alpha^{(m_Q)} \beta^{(m_Q)} (\vec q) \vert H_{(\alpha \beta )}^{q \bar q} 
\vert \psi^{i~(m_Q)}_0  \rangle \vert^2}
\over
{E_i^{0~(m_Q)}-(\epsilon^{(m_Q)}_{\alpha}+\epsilon^{(m_Q)}_{\beta}+{q^2 \over {2\mu_{\alpha \beta}}})}
}
\end{equation}
where the superscripts $(m_Q)$ denote quantities at finite $m_Q$ 
in contrast to those previously defined for $m_Q \rightarrow \infty$.

   I begin by examining the limit \cite{versionone}
$\epsilon^{(m_Q)}_{\alpha}+\epsilon^{(m_Q)}_{\beta}>>E_i^{0~(m_Q)}$ 
and ${q^2 \over {2\mu_{\alpha \beta}}}$, each of which are
for $m_Q$ comparable to $\Lambda_{QCD}$ themselves of order $\Lambda_{QCD}$. In this limit we have
\begin{equation}
\Delta E_i^{dynamic(\alpha \beta)(m_Q)}
~~~~ 
{\buildrel \epsilon^{(m_Q)}_{\alpha}+\epsilon^{(m_Q)}_{\beta} \rightarrow \infty \over \longrightarrow}
~~~~
\langle \psi^{i~(m_Q)}_0 \vert \Delta \tilde V_{\alpha \beta }^{(m_Q)} \vert \psi^{i~(m_Q)}_0  \rangle 
\end{equation}
where
\begin{equation}
\Delta \tilde V_{\alpha \beta }^{(m_Q)}
=
{-1 \over \epsilon^{(m_Q)}_{\alpha}+\epsilon^{(m_Q)}_{\beta}}
\int d^3q 
H_{(\alpha \beta )}^{q \bar q} \vert \alpha^{(m_Q)} \beta^{(m_Q)} (\vec q) \rangle 
\langle \alpha^{(m_Q)} \beta^{(m_Q)} (\vec q) \vert H_{(\alpha \beta )}^{q \bar q}
\label{eq:Vimproved}
\end{equation}
is an $m_Q$-dependent but $\vert \psi^{i~(m_Q)}_0 \rangle$-independent effective 
potential operator. Thus
\bigskip
\begin{eqnarray}
\langle Q \bar Q (\vec r~') \vert \Delta \tilde V_{\alpha \beta }^{(m_Q)} \vert Q \bar Q (\vec r) \rangle
&=&
{-1 \over \epsilon^{(m_Q)}_{\alpha}+\epsilon^{(m_Q)}_{\beta}}
\int d^3q 
\langle Q \bar Q(\vec r~') \vert H_{(\alpha \beta )}^{q \bar q} \vert \alpha^{(m_Q)} \beta^{(m_Q)} (\vec q) \rangle
 \nonumber \\ && 
~~~~~~~~~~~~~~~~~~~~~~\langle \alpha^{(m_Q)} \beta^{(m_Q)} (\vec q) 
\vert H_{(\alpha \beta )}^{q \bar q} \vert Q \bar Q (\vec r) \rangle
\\  \nonumber \\
&=&  
{-1 \over \epsilon^{(m_Q)}_{\alpha}+\epsilon^{(m_Q)}_{\beta}}
\int d^3q ~d^3\rho ~' ~d^3\rho~
\langle Q \bar Q (\vec r~') \vert H_{(\alpha \beta )}^{q \bar q} \vert \alpha^{(m_Q)}\beta^{(m_Q)}(\vec \rho ~')\rangle 
 \nonumber \\
&&~~~~~~~~~~~~~~~~~~~~~~{{e^{i \vec q \cdot (\vec \rho~'-\vec \rho)}
}
\over
{(2 \pi)^3 }
}
\langle \alpha^{(m_Q)} \beta^{(m_Q)} (\vec \rho ) \vert H_{(\alpha \beta )}^{q \bar q} \vert  Q \bar Q (\vec r) \rangle 
\\ \nonumber \\
&=& {-1 \over \epsilon^{(m_Q)}_{\alpha}+\epsilon^{(m_Q)}_{\beta}}
  \int  ~d^3\rho~
\langle Q \bar Q (\vec r~') \vert H_{(\alpha \beta )}^{q \bar q} \vert \alpha^{(m_Q)} \beta^{(m_Q)} (\vec \rho )  \rangle 
 \nonumber \\
&& 
~~~~~~~~~~~~~~~~~~~~~~\langle \alpha^{(m_Q)} \beta^{(m_Q)} (\vec \rho ) 
\vert H_{(\alpha \beta )}^{q \bar q} \vert  Q \bar Q (\vec r) \rangle 
\label{eq:Vimproved}
\end{eqnarray}
\vskip 0.5cm
\noindent which can be compared to Eq. (\ref{eq:DeltaVadiabatic}). I next introduce the finite $m_Q$ analogue of
Eq. (\ref{eq:defc}). As $m_Q \rightarrow \infty$, the form of Eq. (\ref{eq:defc})
is model independent, with dynamical effects residing in $c_{\alpha \beta}(\vec r)$. However, 
for finite $m_Q$ even the form of the analogue to Eq. (\ref{eq:defc}) becomes model-dependent. 
The key to extending the utility of the heavy quarkonium framework down to light quark masses
is to make a ``local approximation"
\begin{equation}
\langle \alpha^{(m_Q)} \beta^{(m_Q)} (\vec \rho ) \vert H_{pc}^{q \bar q} \vert  Q \bar Q (\vec r) \rangle 
\simeq
c^{(m_Q)}_{\alpha \beta}( \vec r) \delta^3(\vec \rho - \eta \vec r) 
\label{eq:defc(m)} 
\end{equation}
where as $m_Q \rightarrow \infty$, $\eta \rightarrow 1$ and
\begin{equation}
c^{(m_Q)}_{\alpha \beta} (\vec r)
\rightarrow c_{\alpha \beta}(\vec r)~,
\end{equation}
the right hand side being the function defined in the adiabatic limit by Eq.~(\ref{eq:defc}).
Note that $c^{(m_Q)}_{\alpha \beta}( \vec r)$ involves at the microscopic level overlap
integrals between $\vert Q \bar Q(\vec r) \rangle$ and
$\vert \alpha^{(m_Q)} \beta^{(m_Q)} (\vec \rho) \rangle$ with wavefunctions
$\psi^{(m_Q)}_{\alpha}(\vec r_{\bar q Q})$ and $\psi^{(m_Q)}_{\beta}(\vec r_{q \bar Q})$
for finite $m_Q$, while $c_{\alpha \beta}( \vec r)$ involves the heavy quark limits
$\psi_{\alpha}(\vec r_{\bar q Q})$ and $\psi_{\beta}(\vec r_{q \bar Q})$ of these wave 
functions. In the simplest and most common models \cite{GIonV,KI,3P0},
the ``local approximation" is exact and automatic with $\eta={m_Q \over {m_Q+m_q}}$, corresponding to
$q \bar q$ pair creation that is point-like and instantaneous. There are, of course,
other possibilities, both local and nonlocal \cite{paircreation}; the latter 
need to be ``localized" {\it via} an approximation
of the form of Eq. (\ref{eq:defc(m)}) in order that one may define their improved
effective potentials.

   With (\ref{eq:defc(m)}) used in (\ref{eq:Vimproved}), we have
\begin{eqnarray}
\langle Q \bar Q (\vec r~') \vert \Delta \tilde V_{\alpha \beta }^{(m_Q)} \vert Q \bar Q (\vec r) \rangle
&=& \delta^3(\vec r~'-\vec r)
\Bigl[ 
{{ - \eta^{-3} \vert c^{(m_Q)}_{\alpha \beta}( \eta \vec r) \vert^2} \over 
{\epsilon^{(m_Q)}_{\alpha}+\epsilon^{(m_Q)}_{\beta}}} \Bigr]
 \\
& \equiv &\delta^3(\vec r~'-\vec r)
\Delta \tilde V^{improved}_{\alpha \beta}(\vec r)~.
\end{eqnarray}
By construction, $\Delta \tilde V^{improved}_{\alpha \beta}$
gives a $\Delta E_i^{improved(\alpha \beta)}$ which converges to
$\Delta E_i^{dynamic(\alpha \beta)(m_Q)}$ for models in which the local approximation
(\ref{eq:defc(m)}) is exact. In all cases \cite{paircreation},
once the local approximation (\ref{eq:defc(m)}) is made,
$\Delta \tilde V^{improved}_{\alpha \beta}$ will give
a more accurate approximation
to the high energy effects of $H_{pc}^{q \bar q}$ than $\Delta V^{adiabatic}_{\alpha \beta}(\vec r)$, so
its use leads to a more rapidly converging approximation to the effects of thresholds on the hadronic system.
We can actually improve matters even further if we incorporate the additional
convergence available even for low mass thresholds as $m_Q \rightarrow \infty$ by defining \cite{versionone}
\begin{equation}
\Delta  V^{improved}_{\alpha \beta}(\vec r) \equiv 
\Bigl[ 
{{  \eta^{-3} \vert c^{(m_Q)}_{\alpha \beta}( \eta \vec r) \vert^2} \over 
{br - (\epsilon^{(m_Q)}_{\alpha}+\epsilon^{(m_Q)}_{\beta})}}\Bigr]~.
\label{eq:finalresult}
\end{equation}
For $\epsilon^{(m_Q)}_{\alpha}+\epsilon^{(m_Q)}_{\beta}$ large, $br$ is negligible so this
approximation is no worse than that leading to $\Delta \tilde V^{improved}_{\alpha \beta}(\vec r)$,
but $\Delta  V^{improved}_{\alpha \beta}(\vec r)$ 
also approaches our old $\Delta V^{adiabatic}_{\alpha \beta}(\vec r)$ as $m_Q \rightarrow \infty$ and so
gives a good approximation to $\Delta E_i^{dynamic(\alpha \beta)(m_Q)}$ for all $\alpha \beta$
in this limit.

   Given these features, we can improve upon Eqs.~(\ref{eq:H}) and (\ref{eq:Hmatrix}) by defining
\begin{equation}
H_{improved}={p^2 \over 2\mu_{Q \bar Q}}+V^{improved}_{n_f}
\end{equation}
and
\begin{equation}
  H^{(\alpha \beta )}_{improved} = 
  \left[ \matrix{
  {p^2 \over 2\mu_{Q \bar Q}}+
   V^{improved}_{n_f} -\Delta V^{improved}_{\alpha \beta }  & H_{(\alpha \beta )}^{q \bar q} & \cr
   H_{(\alpha \beta )}^{q \bar q}                            & {p^2_{\rho} \over 2\mu_{\alpha \beta}}+\epsilon^{(m_Q)}_{\alpha}+\epsilon^{(m_Q)}_{\beta}& \cr
    }\right]
\end{equation}
\vskip 0.5cm
\noindent along with the analogue of Eq.~(\ref{eq:DeltaE})
\begin{equation}
\delta E_i^{(\alpha \beta)}
\equiv
-\langle \psi^{i~(m_Q)}_0 \vert \Delta V^{improved}_{\alpha \beta} \vert \psi^{i~(m_Q)}_0 \rangle
+
\Delta E_i^{dynamic(\alpha \beta)(m_Q)}~.
\end{equation}
The $\delta E_i^{(\alpha \beta)}$ now approach zero both in the strict
adiabatic limit $m_Q \rightarrow \infty$ for all $\alpha \beta$ and also in the limit
$\epsilon^{(m_Q)}_{\alpha}+\epsilon^{(m_Q)}_{\beta}  >> \Lambda_{QCD}$ for all $m_Q$.
They therefore allow a systematic {\it low energy} expansion of the impact
of thresholds on the spectra of {\it all} quarkonia.

\bigskip
\section {Conclusions}

    I have presented here a formalism for calculating the nonadiabatic component 
$\Delta E_i^{\alpha \beta}$ of the mass shift of a
valence heavy $Q \bar Q$ state $i$ from the hadronic loop process $i \rightarrow {\alpha \beta } \rightarrow i$, 
{\it i.e.}, the component of this process that cannot be
absorbed into the renormalized heavy quarkonium potential.  The resulting formula was shown to have the 
expected property
that 
$\Delta E_i^{(\alpha \beta )} \rightarrow 0$
as $m_Q \rightarrow \infty$.    
The formula is also very simple and, when combined with a pair creation model like the flux-tube-breaking model
\cite{KI} or the $^3 P_0$ model \cite{3P0}, should provide a quick method of estimating the influence of nearby 
thresholds on the spectra of heavy quarkonia.

   I have also shown how to define an ``improved" quarkonium potential
which incorporates nonadiabatic effects associated with high mass thresholds for any $m_Q$. When this
potential is identified with the empirical quark model potential, the deviations $\delta E_i^{(\alpha \beta)}$
of the spectrum from the potential model predictions due to thresholds have the property that they
vanish both as  $m_Q \rightarrow \infty$ for all $\alpha \beta$ and also as the mass
$\epsilon^{(m_Q)}_{\alpha}+\epsilon^{(m_Q)}_{\beta}$ of the threshold $\alpha \beta$ gets large for any $m_Q$. 
This improved potential therefore allows a systematic low energy expansion of the
impact of thresholds on hadronic spectra.

   This advantage has a price: the ``improved" potential has the  characteristic that it violates the rule
of flavor independence. While this rule is valid in the heavy quark limit and to leading order in  perturbative QCD
for light quarks, violations are to be expected. Indeed, though obscured by
possible relativistic corrections, there are indications from quark models that the best empirical potentials
are system-dependent \cite{baryonvsmeson}. 

   An important step not taken here is to
calculate the $\Delta E_i^{(\alpha \beta)}$ and $\delta E_i^{(\alpha \beta)}$ for selected channels
to assess numerically how rapidly each converges as 
$\epsilon^{(m_Q)}_{\alpha}+\epsilon^{(m_Q)}_{\beta} \rightarrow \infty$ \cite{versionone},
and to quantify the $m_Q$-dependence of $V^{improved}_{n_f}$. Quark models seem to
constrain this mass dependence to be surprisingly weak \cite{GodfreyIsgur}. 
Assuming that the approach defined here passes quantitative tests such as these, it will then be interesting to
apply it to a number of outstanding phenomenological issues. Among these are the threshold 
shifts in the $c \bar c$ and $b \bar b$ systems and the $\Lambda(1520)-\Lambda(1405)$ problem.
It will also be amusing to study heavy-light systems to see explicitly how groups of states conspire to 
maintain the spectroscopic relations required by heavy quark symmetry 
\cite{IWspec} as $m_Q \rightarrow \infty$, and to quantify the importance of 
heavy-quark-symmetry-breaking pair creation effects residing in the $\delta E_i^{(\alpha \beta)}$
compared to their valence potential model counterparts \cite{NIonLS}.

     Finally, I note that while this paper in couched in the language of the nonrelativistic quark model,
there is nothing in the proposed general framework that would prevent its being transferred to either a relativistic
quark model or to field theory. 

\bigskip

\vfill\eject

{\centerline {\bf REFERENCES}}


\begin{references}
\bibitem{Zcomment} Lowest order meson loop Feynman graphs can have two distinct topologies at the
quark level:  those with a closed $q \bar q$ loop and those where only the (relativistically) 
propagating valence lines appear.  Both types of graphs, when analyzed in a hadronic basis, have mesons
``in the air".  In mesons, the latter class of graphs either correspond to $\eta'$ meson loops 
(and so are dual to $Z$ graphs) or to direct ``s-channel" $\omega-\phi$-mixing-type
OZI violations \cite {GIonOZI}.  In baryons, such graphs correspond to a variety of meson loops, but
all are dual to Z-graphs.  Given these dualities, one can imagine that
non-closed-$q \bar q$-loop diagrams are
subsumed into the valence quark model.  We focus here on the pure $q \bar q$ loop processes which are not
otherwise taken into account in the quark model.
\bibitem{Q1Q2} The considerations of this paper are trivially extended to mesons $Q_1 \bar Q_2$ with 
arbitrary masses $m_1$ and $m_2$.
\bibitem{versionone} One could instead expand Eq. (\ref{eq:DeltaEdynamic}) in 
$\epsilon^{(m_Q)}_{\alpha}+\epsilon^{(m_Q)}_{\beta}+{q^2 / {2\mu_{\alpha \beta}}}$.
This would lead to a Yukawa-like nonlocality in $\vert \vec \rho~'- \vec \rho \vert$
which, using a local approximation to this nonlocality and an analogue to Eq. (\ref{eq:defc(m)}),
would give an alternative definition of $\Delta V^{improved}_{\alpha \beta}(\vec r)$.
This version of $\Delta V^{improved}_{\alpha \beta}(\vec r)$ would not contain the
adiabatic limit as $m_Q \rightarrow \infty$, and so probably
converges more slowly than Eq. (\ref{eq:finalresult}).
\bibitem{paircreation} If the pair creation occurs by creating a gap of classical length $2m_d/b$ in the
flux tube, then $\eta={m_Q \over {m_Q+m_q}}+{2m_q^2 \over {(m_Q+m_q) br}}$. In other models the pair creation might
be intrinsically nonlocal (for an example, see Ref. \cite{GIonV}); in such cases
the definition of $\Delta V^{improved}_{\alpha \beta}$ requires that one make a local
approximation of the form of Eq. (\ref{eq:defc(m)}) to the model's actual matrix element 
$\langle \alpha^{(m_Q)} \beta^{(m_Q)} (\vec \rho ) \vert H_{pc}^{q \bar q} \vert  Q \bar Q (\vec r) \rangle$.
\end{references}
\end{document}